\documentclass[manuscript]{aastex631}
\usepackage{epstopdf}
\usepackage{tikz}
\usepackage{bm}
\usepackage{verbatim}

\newcommand\aastex{AAS\TeX}

\received{xx}
\revised{xx}
\accepted{xx}
\submitjournal{PSJ}

\shorttitle{\aastex\ sample article}
\shortauthors{Cai, Chan, \& Mayr}

\begin{document}

\title{Deep, Closely-Packed, Long-Lived Cyclones on Jupiter's Poles}

\correspondingauthor{Kwing L. Chan}
\email{klchan@must.edu.mo}

\author[0000-0003-3431-8570]{Tao Cai}
\affil{State Key Laboratory of Lunar and Planetary Sciences, Macau University of Science and Technology, Macau, People's Republic of China}

\author[0000-0002-6428-1812]{Kwing L. Chan}
\affil{State Key Laboratory of Lunar and Planetary Sciences, Macau University of Science and Technology, Macau, People's Republic of China}

\author{Hans G. Mayr}
\affil{Emeritus, NASA Goddard Space Flight Center, Greenbelt, USA}



\begin{abstract}
Juno Mission to Jupiter has found closely-packed cyclones at the planet's two poles. The observation that these cyclones coexist in very confined space, with outer rims almost touching each other but without merging, poses a big puzzle. In this work, we present numerical calculations showing that convectively sustained, closely-packed cyclones can form and survive without merging for a very long time in polar region of a deep rotating convection zone (for thousands of planetary rotation periods). Through an idealized application of the inertial stability criterion for axisymmetric circulations, it is found that the large Coriolis parameter near the pole plays a crucial role in allowing the cyclones to be packed closely.
\end{abstract}




\section{Introduction}\label{sec:intro}
NASA's Juno spacecraft has discovered closely-packed cyclones on both the North and South polar regions of Jupiter \citep{bolton2017jupiter,adriani2018clusters,tabataba2020long}. The Northern Polar Cyclone, about 4000 km in diameter, is surrounded by eight circumpolar cyclones with similar sizes. The Southern Polar Cyclone was surrounded by five circumpolar cyclones, with diameters ranging from 5600 km to 7000 km, but recent observations have found that the number of surrounding circumpolar cyclones increased to six and then changed back to five again \citep{adriani2020two}. Aside from the presence of small precessions around the poles, the cyclones are quite stationary. Measured at 1500 km away from their respective centers, the cyclonic rotation periods range from 27.5 to 60 hours \citep{adriani2018clusters}. Stationary cyclonic vortices have also been observed on Saturn's polar regions, but only one strong cyclone dominates at each pole \citep{godfrey1988hexagonal,vasavada2006cassini}.

It is well known that beta effect associated with rotation of a sphere can cause cyclones to migrate poleward and anticyclones to migrate equatorward \citep{schecter1999vortex}. Accumulation of cyclonic vorticity in polar region has been demonstrated by potential vorticity calculation based on contour dynamics \citep{scott2011polar}. Idealized aqua-planet general circulation model experiments have shown that tropical cyclones tend to cluster toward the pole \citep{merlis2016surface}. For giant planets, polar cyclonic vortices have been examined by a shallow-water model in the setting of $\gamma$-plane \citep{o2015polar,nof1990modons,o2016weak} (the Coriolis parameter is a quadratic function of distance to the pole, also called polar $\beta$-plane). With Saturn's parameters, the model, driven by hypothesized moist convection in the surface weather layer, created a strong polar cyclonic vortex. For Jupiter, the model predicted random vortices near the pole \citep{o2016weak}. Using a different shallow $\gamma$-plane model, \citet{brueshaber2019dynamical} investigated the effects of randomly injected storms (local mass perturbations) on polar flows. They found that a large cyclonic polar vortex is usually formed when the Burger number (the square of the ratio of Rossby deformation radius to planetary radius) is large. Multiple small vortices can be formed when this number is small. However, the simulated cyclones do not remain in the pole for significant length of time.

In this paper we consider a dynamically self-consistent approach based on solving the compressible flow equations for a deep rotating convection zone (CZ). Juno's gravitational measurements indicate that Jupiter's zonal winds extend 3000 kilometers below the visible surface at low latitudes, and $0-500$ kilometers at high latitudes \citep{kaspi2018jupiter}. Despite the high uncertainty in the polar region, it does not contradict with the presence of a deep CZ. Numerical simulations \citep{chan2007rotating,kapyla2011starspots,chan2013numerical,guervilly2014large,stellmach2014approaching,favier2014inverse,aurnou2015rotating,cai2016semi,heimpel2016simulation,guervilly2017jets,yadav2020deep} have demonstrated that large-scale, long-lived vortices (cyclonic or anticyclonic) can be generated spontaneously in a fast rotating CZ. Our numerical model, to be discussed in Section~\ref{sec:model}, is along this pathway.

Simulation of vortices in a shearing zonal flow indicated that two stable vortices tend to merge if their cross-zonal distance is smaller than a critical value approximately equal to the vortices' radii \citep{marcus1990vortex}. Two same-sign vortices generally merge when the separation is below a certain threshold value (also of the order of the vortex radius) \citep{cerretelli2003physical,folz2017quantitative}. A principal puzzle, therefore, is why Jupiter's closely-packed polar cyclonic vortices do not merge. In this study, we provide numerical evidence that convection generated long-lived cyclones in the polar $\gamma$-box (three-dimensional) can be packed without merging for a long time, and an idealistic application of the inertial stability criterion \citep{yanai1964formation,schubert1982inertial,kloosterziel2007inertial} is invoked to provide some insights (Section~\ref{sec:result}).

While considering vortex packing in the polar regions of Jupiter as deep convective phenomenon, one cannot avoid the question of why Saturn displays only a single cyclone in each pole.  We address this question in Section~\ref{sec:discussion}.

\section{The Model}\label{sec:model}
Numerical simulations are performed by solving the fully compressible hydrodynamic equations for an ideal gas in a rectangular box:
\begin{eqnarray}
&&\partial_{t} \rho=-\bm{\nabla}\bm{\cdot} (\rho \bm{v})~,\label{eq1}\\
&&\partial_{t}(\rho \bm{v})=-\bm{\nabla}\bm{\cdot} (\rho \bm{vv})-\bm{\nabla} p +\bm{\nabla}\bm{\cdot}\bm{\Sigma}+\rho \bm{g}+2\rho \bm{v}\bm{\times} \bm{\Omega}~,\\
&&\partial_{t}E=-\bm{\nabla} \bm{\cdot} [(E+p)\bm{v}-\bm{v}\bm{\cdot}\bm{\Sigma}+\bm{F_{\rm d}}]+\rho \bm{v}\bm{\cdot}\bm{g}-\rho c_{p} (T-T_{top})/\tau~,\label{eq3}
\end{eqnarray}
where $\rho$ is the density, $\bm{v}$ is the velocity, $p$ is the pressure, $\bm{\Sigma}$ is the viscous stress tensor, $\bm{g}$ is the gravitational acceleration, $\bm{\Omega}$ is the angular velocity,  $E=e+\rho\mathbf{\it{v}}^\mathbf{2}/2$ is the total energy density, $e$ is the internal energy, $\bm{F_{\rm d}}$ is the diffusive energy flux, $T$ is the temperature, $c_{p}$ is the heat capacity under constant pressure, $\tau$ is the time scale of Newton cooling (to mimic fast radiative loss above the planet's optical surface), the subscript $top$ denotes the value at the top of the box. An ideal gas law $p=\rho R_{s} T$, where $R_{s}$ is the specific gas constant, is used for the equation of state. The ratio of specific heats of the gas is $\gamma_{ad}=1.47$ \citep{horedt2004polytropes}. We adopt a `large eddy simulation' approach in which sub-grid-scale (SGS) turbulent processes in the momentum equation are modeled by a SGS kinematic viscosity $\nu$ according to the Smagorinsky scheme \citep{smagorinsky1963general}
\begin{eqnarray}
\nu=c_{\nu}^2 \Delta x \Delta z(2\bm{\sigma}:\bm{\sigma})^{1/2}~,
\end{eqnarray}
where $\Delta x$, $\Delta z$ are local sizes of the horizontal and vertical grids, $\bm{\sigma}$ is the strain rate tensor, the double dot denotes tensor contraction, and the coefficient $c_\nu$ is chosen to be 0.28.

The hydrodynamic equations (1-3) are solved by an explicit finite-difference method. The numerical code is written in a way that conserves total mass, total energy, and momentum to round-off limits. The computational domain is discretized by a staggered grid. Pressure, density, and temperature are located on grid levels, while velocities are located on half-grid levels. In the original version of this code \citep{chan1986turbulent}, the equations were integrated by a semi-implicit method that can accommodate time steps a few times larger than the Courant-Friedrichs-Lewy restriction. For the simulations performed in this paper, the Mach number is on the order of $O(0.1)$, and the speed-up achieved by the semi-implicit treatment of time integration is not advantageous versus efficiency of parallelization. For this reason, we instead use an explicit predictor-corrector method.

The three-dimensional numerical model, apart from the location-dependent Coriolis terms and some differences in parameters, is essentially the same as the one used in \citet{chan2013numerical}. The rectangular `$\gamma$-box' is considered to represent a piece of spherical shell with polar axis through its center, along the vertical direction. The local gravitational acceleration $\bm{g}$ defines the vertical direction ($z$) for all points in the box (it is a constant in the present model). The components of the Coriolis term vary with location as the rotation vector $\bm{\Omega}$ can be tilted with respect to $\bm{g}$. The inclined angle between $\bm{\Omega}$ and $-\bm{g}$ is the colatitude $\theta$, and the latitude is $\phi=90^o-\theta$. The form of the vector equations remain unchanged, but $\bm{\Omega}$ now depends on location. As the box is three-dimensional, the effects of both the vertical Coriolis parameter $f=2\mathrm{\Omega \sin\phi}$ ($=2\mathrm{\Omega \cos\theta}$ in terms of the colatitude) and the horizontal Coriolis parameter $f^\prime=2\mathrm{\Omega \cos\phi}$ ($=2\mathrm{\Omega \sin\theta}$) are included. Here $\mathrm{\Omega}=\|\mathbf{\Omega}\|$, $\xi=((x-x_{c})^2+(y-y_{c})^2)^{1/2}$ and $\theta=\xi/R$ are the distance and the colatitude of an arbitrary point $(x,y,z)$ to the polar axis (center line of the box where $x/x_{c}=y/y_{c}=1$), and $R$ is the outer radius of the sphere. The horizonal Coriolis parameter $f^\prime$ is small near the pole and turns out not to be influential, but it is included for completeness of the three-dimensional equations. In the spherical shell to rectangular box mapping, the ratio $H/R$, where $H$ is the thickness (or height) of the box, is a relevant parameter. For the approximation to be valid, both $H/R$ and $\theta_{max}$ (the colatitude angle spanned by the half-width of the box) need to be much less than 1. We use the term '$\gamma$-box' to emphasize the finite thickness of the domain. If the thickness of the box can be considered vanishingly small compared to the radius of the spherical shell, this approximation reduces to the ordinary $\gamma$-plane approximation ($f=2\Omega-\Omega(\xi/R)^2$ and $f'=0$).

In our calculation, the side boundaries of the box are periodic and the top and bottom boundaries are stress-free. These conditions ensure that aside from the Coriolis force, no unspecified momentum exchange occurs between the fluid layer and the planet. Furthermore, the stress-free boundary condition is more favorable for the formation of large-scale vortical structures \citep{guervilly2017jets,guzman2020competition}. Constant, uniform energy flux $F_{bot}$ is fed at the bottom, and the temperature at the top is held constant and uniform. All quantities are made dimensionless by setting the thickness of the box, the initial temperature, density, and pressure at the top to 1. As a result, velocity is scaled by the isothermal sound speed (square root of the ratio of pressure to density) at the top. Time scale is given by the amount of time for this isothermal sound speed to travel a distance equal to $H$. To obtain time in terms of the rotation period of the planet, one can multiply the dimensionless time with the dimensionless $\Omega/(2\pi)$. Dimensional quantities can be recovered once the values of the scaling quantities at the top level are specified (see Appendix~\ref{appendixa}).

The box contains two layers: a convectively unstable layer in the lower part and a convectively stable layer in the upper part. In the convectively unstable layer, where the mean vertical temperature gradient is almost adiabatic (but slightly superadiabatic), the thermal conductivity $\kappa$ is set to deliver only 50\% of the total energy flux so that convection needs to pick up the rest of the flux. Due to the significant role of convection in energy transfer, this layer is identified as a convection zone (CZ). In the stable layer, the thermal conductivity is raised to a level that all of the energy flux can be fully delivered by a sub-adiabatic temperature gradient. It is to be labeled as a radiation zone (RZ). In addition, to model the fast loss of energy in the optically thin tropospheric layer, a Newtonian cooling term is smoothly introduced to replace the conduction term in the RZ. After thermal relaxation, it effectively creates a constant temperature layer near the top of the box. The change of atmospheric stability (and thermal conductivity) occurs at the interface between the CZ and the RZ. The location of this interface is at 0.95 of the total height of the box for Jupiter-like simulations (Cases A-B, see later discussion), and at 0.75 for Saturn-like simulations (Cases C-E).

The initial structures of the gas layers are polytropic, i.e. $\rho\propto T^n$, where $n$ is the polytropic index (see \citet{hurlburt1984two} for a detailed description of polytropic structure in numerical modeling).  $n$ takes the value 2.128 (equals to the adiabatic value $n_{ad}=1/(\gamma_{ad}-1)$) in the CZ. The value 9 is adopted in the RZ where a large value of polytropic index is used to mimic the rapid change of density relative to temperature. Note that as long as this number is large, the exact value is not important. The final equilibrium state of the RZ is mainly determined by the Newton cooling which enforces an isothermal layer in most of the RZ.

For the Jovian cases, the CZ and the RZ contain about 4.7 and 0.5 pressure scale heights, respectively. Although our simulations consider a fairly deep density-stratified region, the stratification is still much less than that of Jupiter's outer convection zone (where magnetic effects on the hydrodynamics can be considered small). The Jovian CZ may contain more than 10 pressure scale heights, which is prohibitive for current computer resources (see later discussion on thermal relaxation). Here we intend to take the compressibility effects into account as much as computer resources allow.

The box has a square base; $\lambda$ is the aspect ratio (lateral dimension over height). The aspect ratio has an important effect on the formation of multiple vortices. In earlier simulations \citep{kapyla2011starspots,chan2013numerical,guervilly2014large} where aspect ratios were small, a single vortex often filled the box. To produce multiple vortices, a computational box with large aspect ratio is needed; $\lambda$ is chosen to be 16 in our polygon simulations. Investigations on rapidly rotating convection reveal that the convective Rossby number $Ro^{c}$ (root-mean-square velocity in the CZ of the non-rotating case, computed separately, divided by the thickness of the CZ and by the Coriolis parameter $f$) and Reynolds number $Re$ (root-mean-square velocity of the computed flow multiplied by the thickness of the CZ, and divided by the averaged kinematic viscosity) are two key factors determining the formation of large-scale vortices \citep{guervilly2014large}. $Ro^c$ compares the strength of convective driving to that of the Coriolis effect. Numerical simulations show that $Ro^c$ needs to be smaller than a critical value $\sim0.25$ for long-lived cyclones to be generated \citep{chan2013numerical}. Another necessary condition is that the Reynolds number $Re$ should be $\gg 100$ \citep{guervilly2014large}. As a result, simulation of multiple cyclones needs large $\lambda$, high $Re$, and low $Ro^c$. All these conditions require large number of grids and long integration time. In our numerical experiments, we find that polygon patterns could only be formed in high resolution simulations with long periods of time evolution (see Appendix~\ref{appendixb}). Each computed case typically requires millions of CPU core hours (tens of millions of time steps) to complete. In the current stage, it is difficult to explore the parameter space for a full delineation of the pattern formation boundaries. Our main purpose here is to use a couple of cases to demonstrate that large-scale circumpolar vortices can be produced and survive in polygon patterns for a long time in a deep convection zone.

\begin{deluxetable*}{ccccccccccccccccc}[htb!]
\tablecaption{Parameters of simulation cases for closely-packed cyclones\label{table:tab1}}
\tablehead{
 Case & $Ek$ & $Re$ & $Pr$ & $Ro^{c}$ & $F_{bot}$ & $\theta_{max}$ & $H/R$ & $N_{x}\times N_{y}\times N_{z}$ & {Pattern}
}
\startdata
A &   $2.2\times 10^{-5}$   & $1.8\times 10^4$ & 6.2 & 0.052 & 0.0016 & $24^{\circ}$ & 5.24\% & $1200^2\times 101$ & {\it pentagon} \\
B &   $1.5\times 10^{-5}$   & $1.6\times 10^4$ & 4.2 & 0.041 & 0.001 & $12^{\circ}$ & 2.63\% & $1152^2\times 129$  & {\it hexagon} \\
\enddata
\tablecomments{$Ek$ is the Ekman number (viscosity divided by the Coriolis parameter and the square of the thickness of the CZ); $Re$ is the Reynolds number; $Pr$ is the Prandtl number; $Ro^{c}$ is the convective Rossby number; $F_{bot}$ is the total energy flux; $\theta_{max}$ is the maximum colatitude angle spanned by half of the width of the box; $H/R$ is the ratio of the thickness of the computed zone to the planetary radius; $N_x$,$N_y$, and $N_z$ are grid numbers in the $x$-, $y$-, and $z$-directions, respectively; 'Pattern' denotes the polygonal organization of the vortices. Each case has been run for over 38000 units of time (over 3000 planetary rotation periods).}
\end{deluxetable*}

Parameters of our simulation cases are listed in Table~\ref{table:tab1}. For all the cases, we choose ${\Omega}=0.5$. We first found the polygonal clustering of vortices through Case A. The parameters were changed to bracket some of Jupiter's physical parameters with Case B. We can compare our model parameters with the physical parameters of Jupiter by substituting physical values of the scaling quantities into the scaling factors.
 The surface temperature, density, and pressure of Jupiter are about $166$K, $0.167 \rm kg/m^3$, and $10^5 \rm Pa$, respectively. With these as dimensional values of the scaling quantities at the top of the box, the velocity scale is then $\approx 773.8 \rm m/s$. The radius of Jupiter is about $70000$ km. The box thickness of case A, being 0.0524$R$\rm, is about 3600km (the thickness of the stable layer is about 180 km), and the width is about 58000km. The scale for time in this case is about $4740 \rm s$, with which and the rotation period can be found to be approximately $16.5$ hours. For Case B, the box thickness is about 1800km (the thickness of the stable layer is about 90km), and the width is about 29000km. The planetary rotation period is about 8.3 hours; it is not equal but closer to Jupiter's rotation period of about 10 hours \citep{bagenal2007jupiter}.

Both cases deal with deep stratification. Simulation with realistic values of internal heat flux and stratification of Jupiter is extremely difficult. Even for the shallower case $H/R=2.6\,\%$, the stratification spans 10 pressure scale heights. The observed internal heat flux of Jupiter is about $5.4-7.5 \rm W m^{-2}$ \citep{hanel1981albedo,li2018less}. The thermal relaxation time can be estimated by the Kelvin-Helmholtz timescale $t_{KH}=\int e dz/F_{bot}$, which is over $O(10^5)$ years (and over $O(10^8)$ rotation periods). So far no numerical simulation of Jupiter's deep convection zone can get close to such numbers. In practice, larger values of heat flux and smaller stratification are adopted (e.g.\ Jones 2014). In our simulations, the internal heat flux has been amplified by a factor of about 10000 and the stratification is reduced to about 5 pressure scale heights (through reduction of the physical value of gravitational acceleration by the factors 12.2 and 6.15 for Cases A and B, respectively) to speed up the relaxation. The fraction of radiative flux in the CZ is also raised to speed up the relaxation process. Effectively, these modifications increase the Rossby number relative to that of Jupiter. In dimensionless parameter space, our cases are near the upper boundary region of the convective Rossby number where the clustering phenomenon can occur.

To reduce computational cost, the flow fields were first generated in a smaller {\it f}-plane box (aspect ratio $\lambda=4$) with high grid resolution. At the beginning of calculation, a small random perturbation in the velocity field was introduced to initiate the development of convection. When well-defined small vortices were formed, the flow field of the smaller box was then periodically extended to initialize the flow field for the $\lambda=16$ $\gamma$-box. The thermal structure, turbulence, and organized flows evolved in a mutually consistent way.

\section{Results}\label{sec:result}

\subsection{Vortex Configurations}

\begin{figure}[!htbp]
\epsscale{0.95}
\plotone{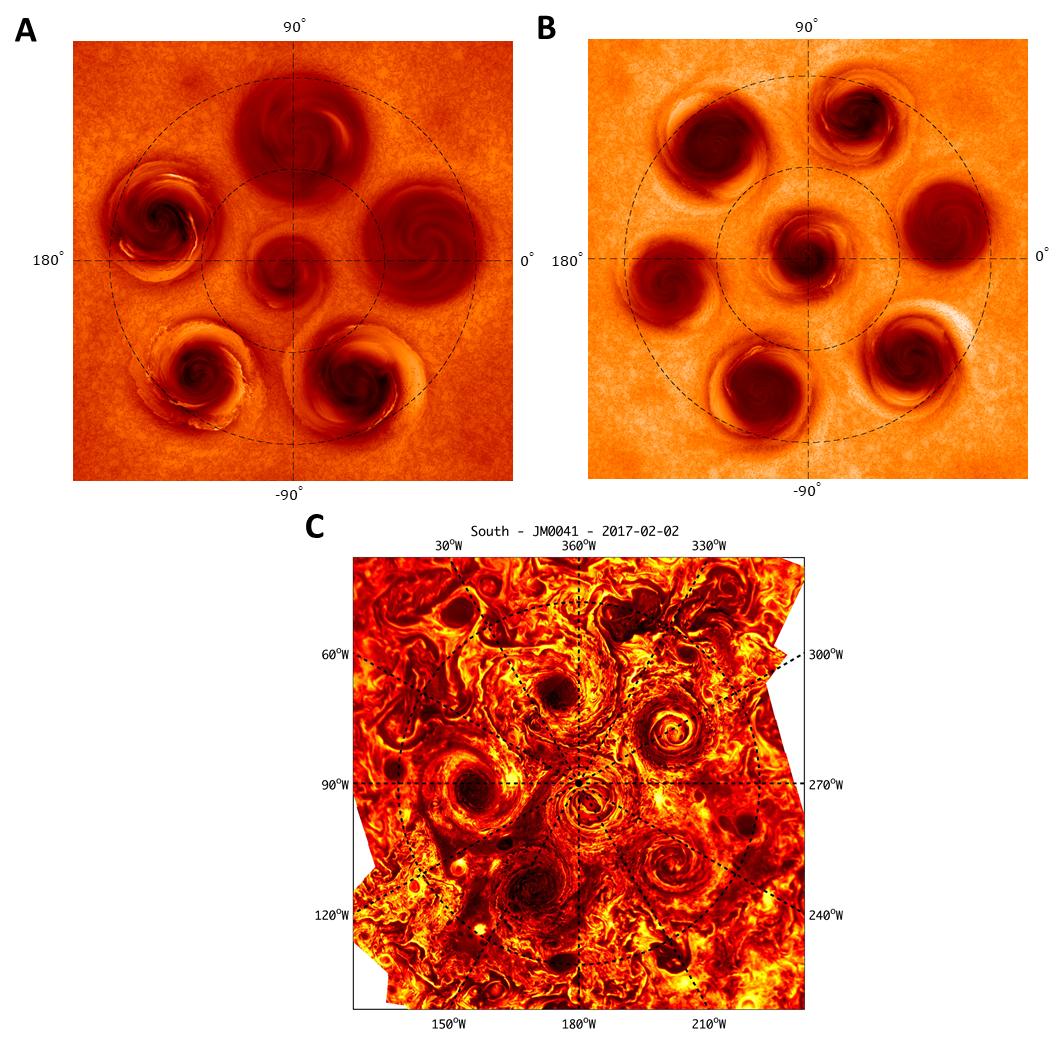}
\caption{Comparison of polygon patterns between simulation and observation. The top panel shows horizontal cuts of instantaneous temperature fields near the radiative-convective boundary: (A) The simulated pentagon case. The large and small dashed circles show the $70^{\circ}$ and $80^{\circ}$ latitudes, respectively. 
In terms of dimensional units, the temperature contrast at this layer is about 35{\rm K};  (B) The simulated hexagon case. The large and small dashed circles show the $80^{\circ}$ and $85^{\circ}$ latitudes, respectively. 
The temperature contrast at this layer is about 22{\rm K}. Spirals at the outer rims of the vortices show up prominently.
The bottom panel (C) shows the observed image by JIRAM at Jupiter's south pole \citep{adriani2018clusters}. Figure1C courtesy of Alessandro Mura and Alberto Adriani. Figure 1C is reprinted by permission from reference \citet{adriani2018clusters}, Springer Nature: Nature, copyright (2018). Note that (A) and (B) are viewed from north (cyclonic motion is counterclockwise), while (C) is viewed from south (cyclonic motion is clockwise). \label{fig:f1}}
\end{figure}

In the polar $\gamma$-box, the cyclonic vortices clustered toward the pole, merged, and grew bigger. Merging was the principal process that facilitated the growth of vortex sizes. After a long period of thermal and dynamical relaxations, the growth process ceased. A stable polygon pattern was finally formed. Figs.~\ref{fig:f1}A and \ref{fig:f1}B show the temperature structures near the radiative-convective boundaries. To illustrate the dynamical processes, time evolution of temperature structures in the small and large boxes for case B are shown in MovieS1 and MovieS2, respectively. The former shows the run in the $\lambda=4$ box (for about 98 rotation periods) where intermediate cyclones first appeared, and the latter shows the extended run in the $\lambda=16$ box (for over 2600 rotation periods). The bright (dark) color represents higher (lower) temperature.

MovieS2 tracks the evolution of the temperature field from the merger generation of size-saturated vortices to the formation of the hexagonal pattern. At the initial stage, small vortices merge to form larger vortices. The merging process essentially stops after the large vortices grow to saturation size. Finally the beta effect around the pole pushes the cyclones in to form a stable hexagonal pattern. From the movie, we also  observed that a nine-vortex pattern, similar to but not as well organized as the octagonal pattern on Jupiter's north polar region, lasted for a while.

The polygon formations of both Cases A and B have been tracked for over 2000 planetary rotation periods. The patterns remain stable. The later portion of MovieS2 already includes a demonstration for Case B.  A shortened demonstration for Case A is given in MovieS3. The movie spans a period of 1200 units of time, corresponding to about 96 planetary rotation periods.

A projected map of Jupiter's south polar region taken at PJ4 from JIRAM \citep{adriani2018clusters} is shown in Fig.~\ref{fig:f1}C for comparison. The cyclones are shaped in a pentagon pattern, with spiral arms at the outer rims of vortices almost touching each other. The pentagon pattern seen from the JIRAM map looks similar to that of Case A. The hexagon pattern that appeared at PJ18 (see the JIRAM image on Figure 13 in \citet{adriani2020two}) looks similar to our Case B shown in Fig.~\ref{fig:f1}B. The diameters of cyclones in Case B are about 6000 {\rm km}, which agree well with the observed values by Juno. For Case A, the central cyclone has similar size, but the diameters of the circumpolar cyclones almost double (This could be related to the larger $Ro^c$, i.e.\ lower rotational constriction, and the larger $\theta_{max}$).

\begin{figure}[!htbp]
\plotone{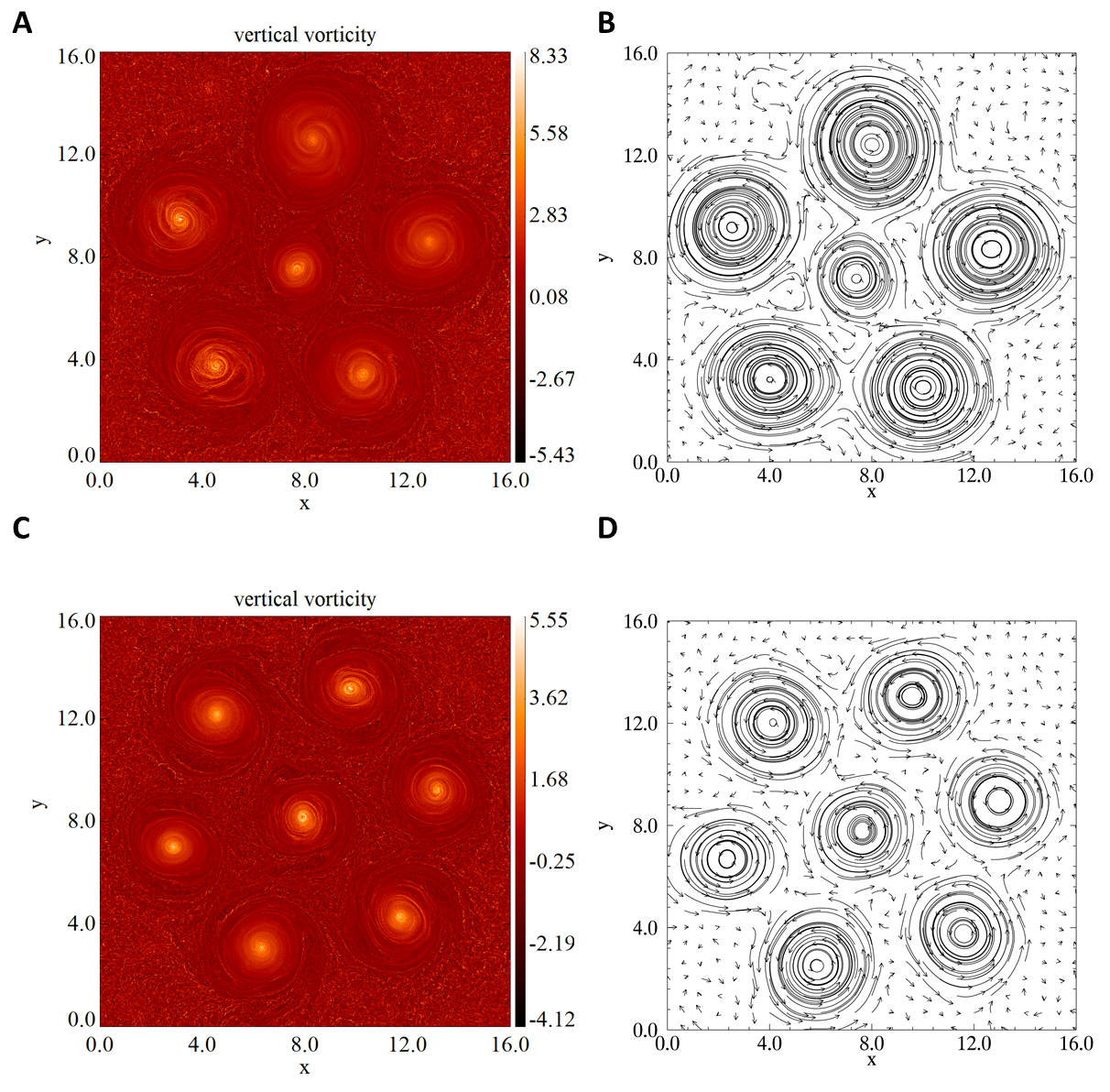}
\caption{Instantaneous flow characteristics in the middle of the box (at $z=0.5$). (A) shows a horizontal cut of the vertical vorticity field of Case A. The bright (dark) color represents higher (lower) vorticity. In dimensional form, one distance unit is about 3668 {\rm km}. (B) shows the corresponding horizontal velocity streamlines.  (C) and (D) are similar to (A) and (B), but for Case B. In dimensional form, one distance unit is about 1840 {\rm km}.  \label{fig:f2}}
\end{figure}

\begin{figure}[!htbp]
\plotone{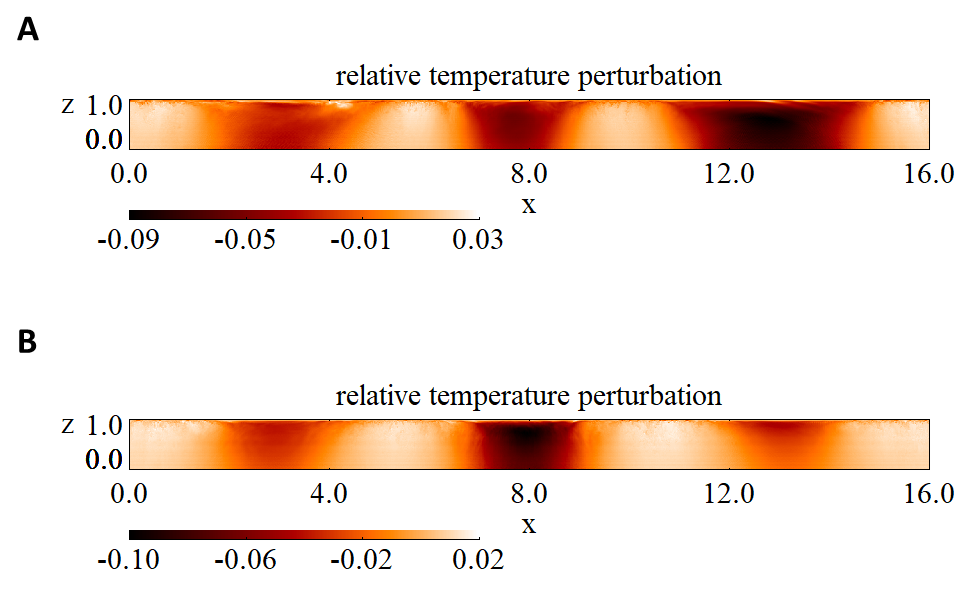}
\caption{Cyclones penetrate the whole depth of the CZ and in a lesser degree extend to the RZ. (A) and (B) show vertical cuts ($y=8$) of the relative temperature variation for the pentagon and hexagon cases, respectively.\label{fig:f3}}
\end{figure}

The vortices have temperature lower than the outside and are cyclonic. Figs.~\ref{fig:f2}A and \ref{fig:f2}C show examples of the vertical vorticity $(\bm{\nabla}\times \bm{v})_z$ cross sections near the top of the CZ and Figs.~\ref{fig:f2}B and \ref{fig:f2}D show the instantaneous streamlines. Positive (cyclonic) vorticity concentrates at centers of the cyclones. The vortices extend through the whole height of the box. This can be observed through Fig.~\ref{fig:f3}, where vertical cuts of relative temperature deviations are shown (the cuts essentially go through the centers of the central vortices but partially miss the outer ones). The temperature deviations drop in the RZ (a thin layer near the top of the box) but the columnar structures are well displayed in the CZ.

While the appearance of different polygons in the two cases illustrates that the cluster patterns need not be unique, the general characteristics bear similarity to those observed by the Juno spacecraft in the southern polar region. The octagon pattern of cyclones in the northern polar region \citep{adriani2018clusters} has not yet been replicated in this paper, but it would not be surprising that a suitable change in some parameters can make that realized. We postpone this for future research.

\begin{figure}[!htbp]
\plotone{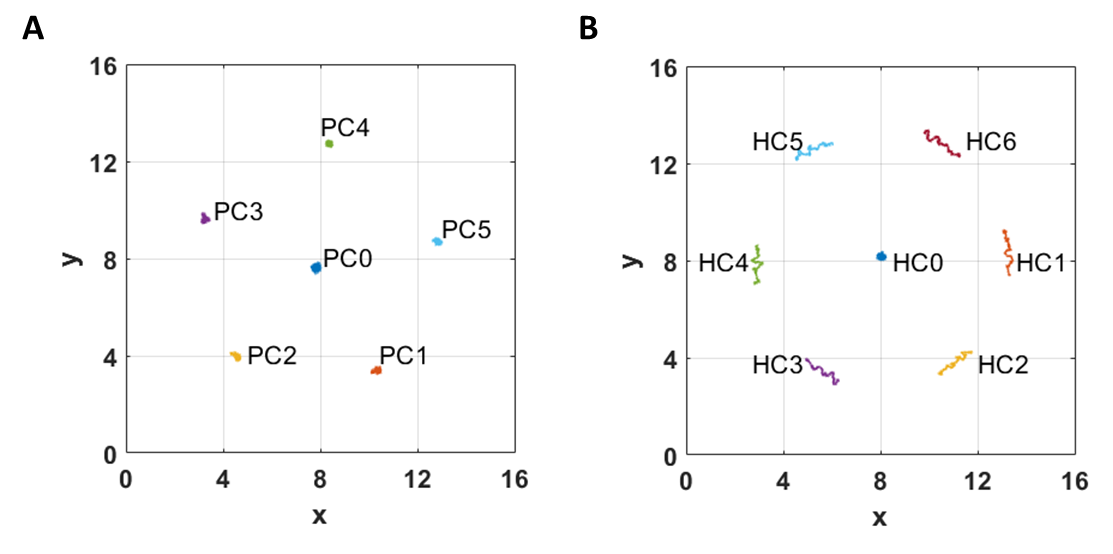}
\caption{Vortex tracks. (A) and (B) track the locations of the centers of the pentagon and hexagon vortices, respectively. The tracking lasted about 96 planetary rotation periods.
\label{fig:f4}}
\end{figure}

\begin{figure}[!htbp]
\plotone{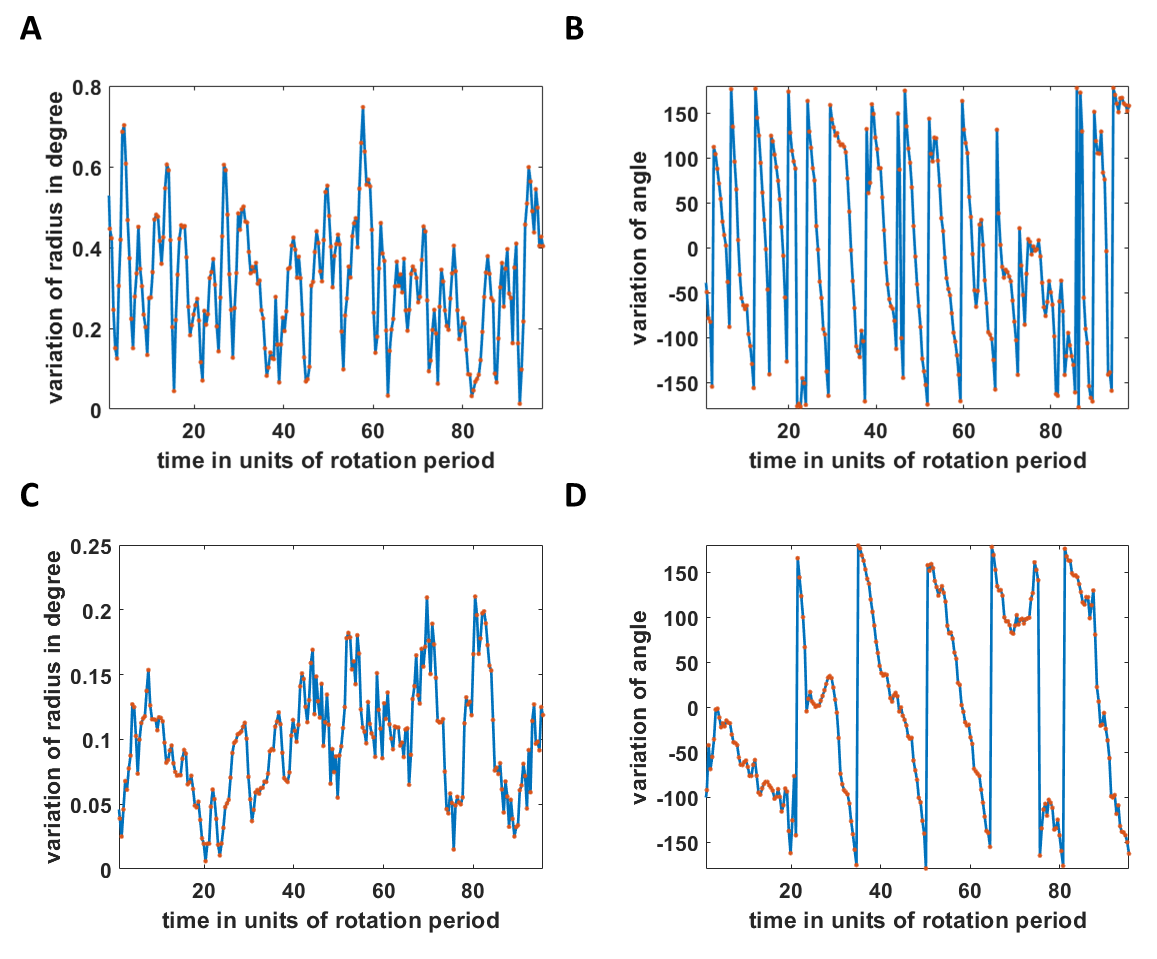}
\caption{Motions of the central vortices. For PC0, (A) shows the instantaneous distance to its time-averaged location and (B) shows the position angle as defined in the main text.  (C) and (D) show corresponding quantities for HC0.
\label{fig:f5}}
\end{figure}

\begin{figure}[!htbp]
\plotone{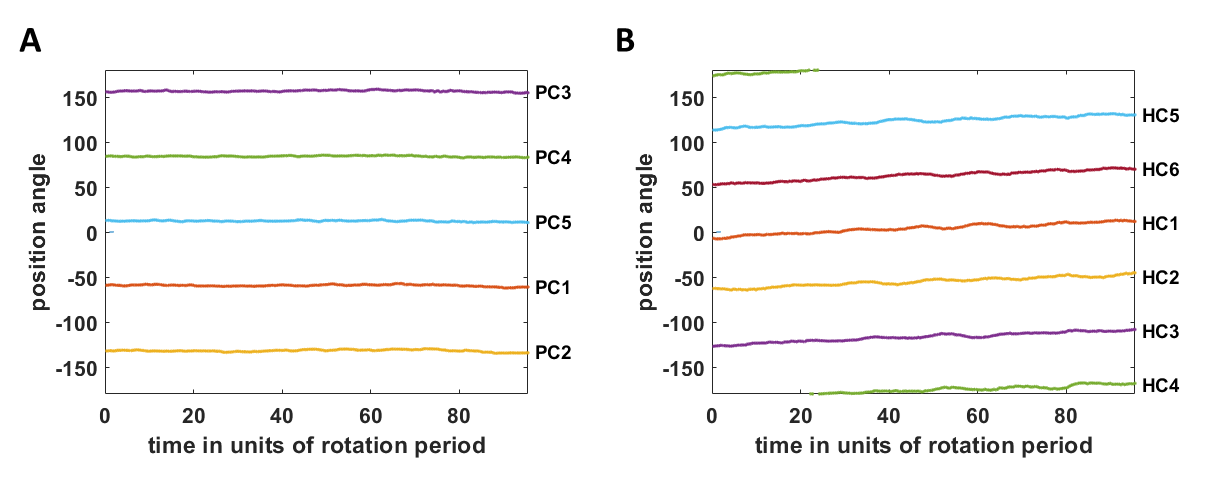}
\caption{Motions of circumpolar vortices. (A) and (B) show position angles of the off-center vortices for the pentagon and hexagon cases, respectively. While the pentagon pattern does not show noticeable rotation, the hexagon pattern drifts around the central cyclone with a rate of about $0.17^{\circ}$ per planetary rotation period.\label{fig:f6}}
\end{figure}

\subsection{Motions of Vortex Centers}

The cyclones are labeled as PC0-PC5 for Case A and HC0-HC6 for Case B. The positions of the centers of the cyclones (locations of local temperature minimum) in the two cases are shown in Fig.~\ref{fig:f4}.
The center of PC0 (the central vortex of Case A) is not exactly at the pole ($x_c=y_c=8$); there is a small offset in the direction away from the largest cyclone PC4 (Fig.~\ref{fig:f4}A). The small offset of the pentagon pattern in Jupiter's south polar region was also observed by Juno \citep{adriani2018clusters}. Despite the stable structure, small regular movements can be detected. One interesting finding is that PC0 drifts around its time averaged center in an anticyclonic fashion, within a distance of $0.75^\circ$.  In Fig.~\ref{fig:f5} the positions of the central cyclones PC0 and HC0 are given in terms of polar coordinates with origins defined to be at their respective time averaged locations; the angle is between the line from the origin to the vortex center and the x-axis (position angle). Fig.~\ref{fig:f5}A shows the temporal variation of the radial distance of PC0 to the origin and Fig.~\ref{fig:f5}B shows its position angle. Figs.~\ref{fig:f5}C and \ref{fig:f5}D show polar coordinates for the central vortex of Case B (HC0). In both cases, the position angle cycles between $-180^o$ and $+180^o$ with negative temporal slopes (anticyclonic motions).

Changes of the relative positions of surrounding cyclones to PC0 are small; the amplitudes of fluctuations in position angles are less than $1.5^\circ$, and there is no obvious systematic drift (Fig.~\ref{fig:f6}A). On the other hand, the hexagonal pattern shows slow cyclonic movement around HC0 (see Figs.~\ref{fig:f4}B and \ref{fig:f6}B; temporal slopes of position angles are positive). HC0 itself drifts around its time averaged center in anticyclonic direction (Fig.~\ref{fig:f5}D). \citet{tabataba2020long} reported that the south circumpolar cyclones also drifted around the central cyclone, with an averaged drift rate of $1.5\pm 0.2^{\circ}$ per Juno's orbital period (about 53 days). Our case B has shown the same sense of drift, but with a high drift rate of about $20^{\circ}$ per Juno's orbital period.

\begin{figure}[!htbp]
\plotone{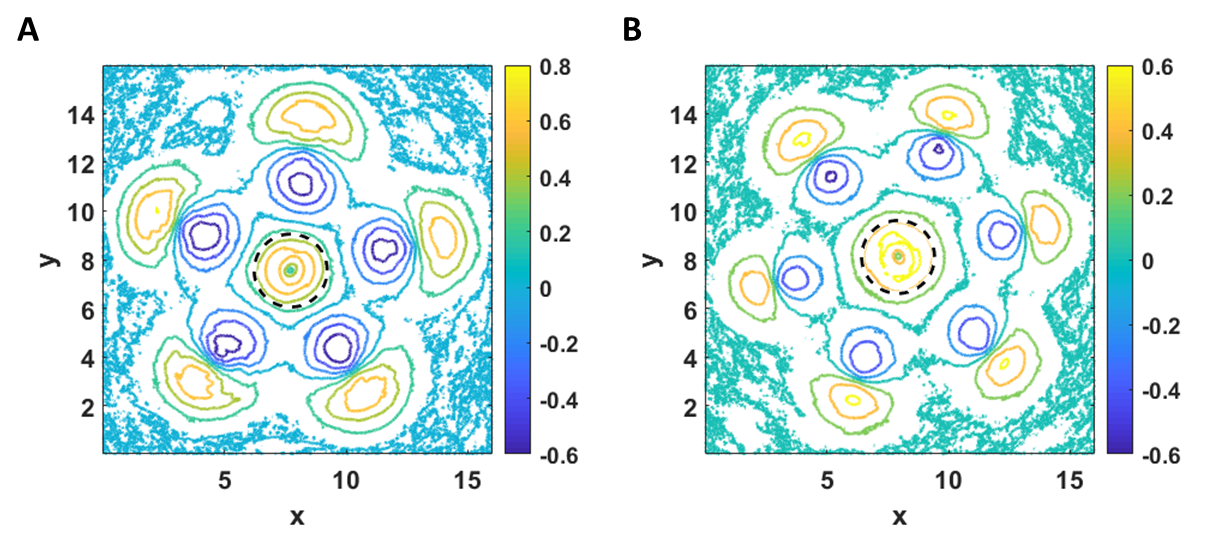}
\caption{Tangential velocity distribution around a vortex.  (A) and (B) are contour plots of the instantaneous tangential velocity (on $z=0.5$) around centers of PC0 and HC0, respectively. The black dashed circles show the distance $r=1.5$.\label{fig:f7}}
\end{figure}

\begin{figure}[!htbp]
\plotone{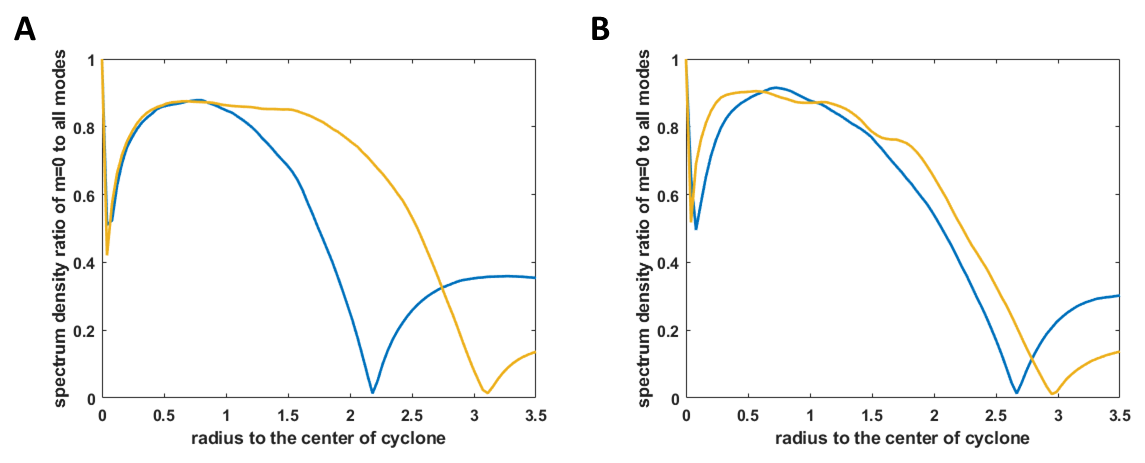}
\caption{The power of the axisymmetric mode $m=0$ relative to the total power in tangential velocity. The values are averaged over 10 planetary rotation periods toward the end of the simulations. (A) plots the ratio for PC0 (blue curve) and PC2 (orange curve); (B) plots the ratio for HC0 (blue curve) and HC2 (orange curve).
\label{fig:f8}}
\end{figure}

\begin{figure}[!htbp]
\plotone{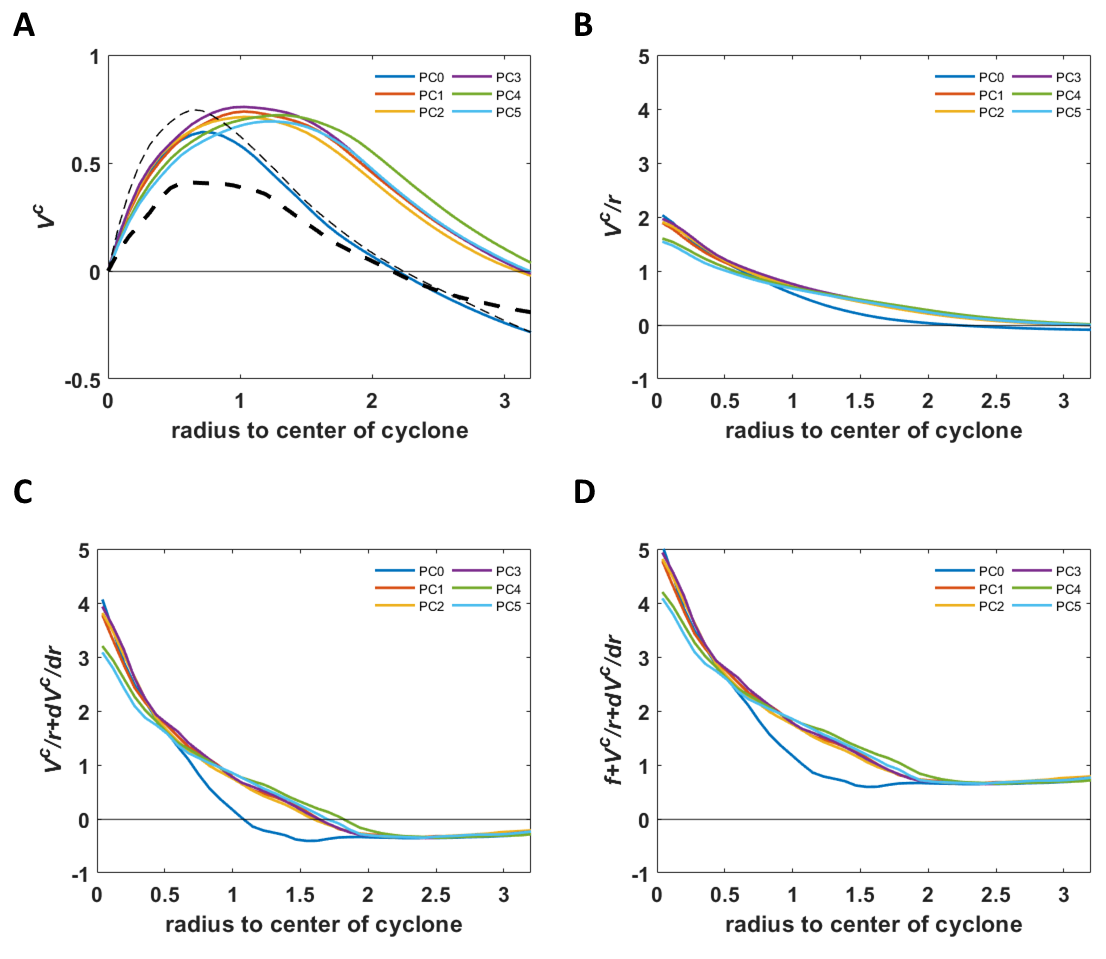}
\caption{Terms affecting stability in the pentagon case. Curves of different colors represent different vortices on cutting plane z=0.5 (except the dashed curves). The values are averaged over 10 planetary rotation periods toward the end of the simulation. (A) Mean tangential velocity $V^c$ around an individual vortex, plotted against radius to the vortex's center. The thin and thick dashed curves are for the cutting planes $z=0.04$ and $z=0.96$, respectively. (B) Mean curvature vorticity $V^c/r$. (C) Relative vorticity $V^c/r+{dV}^c/dr$. (D) Inertial stability factor $f+V^c/r+{dV}^c/dr$. \label{fig:f9}}
\end{figure}

\begin{figure}
\plotone{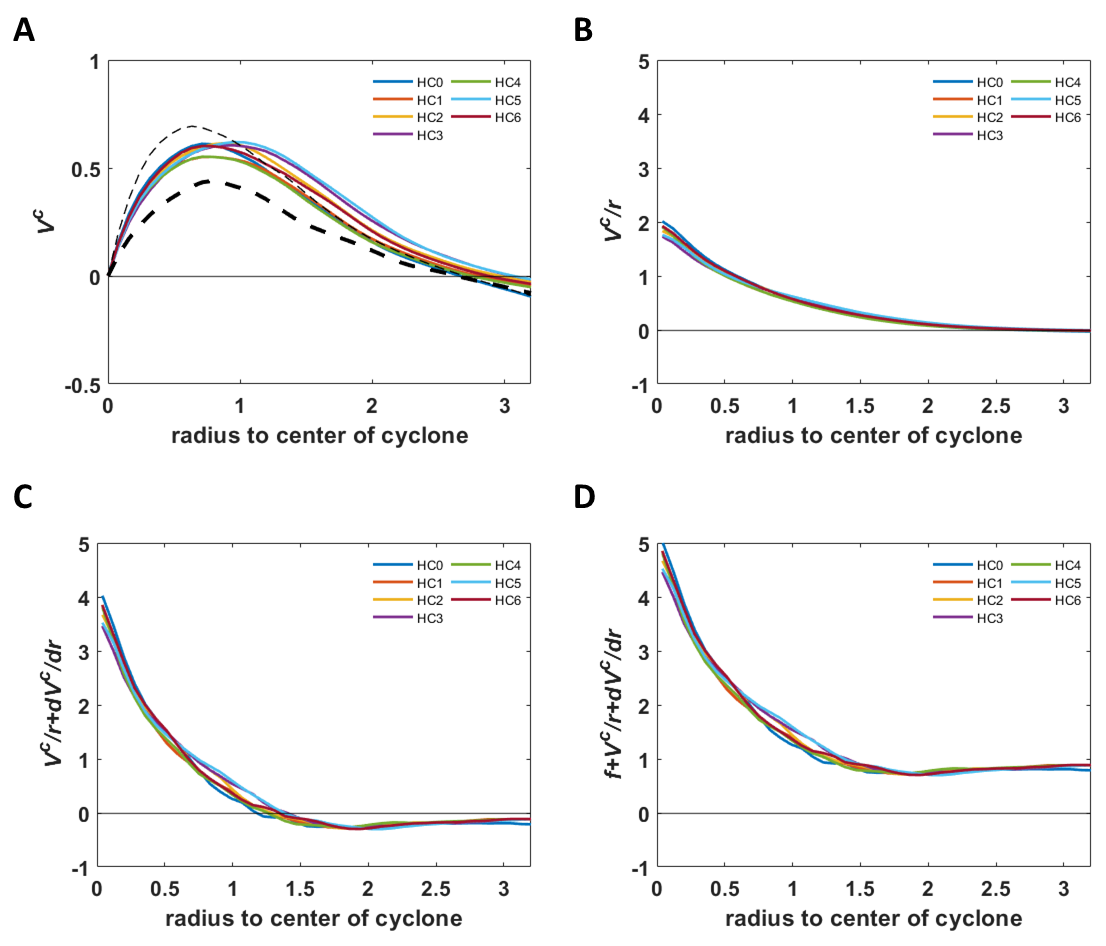}
\caption{Terms affecting stability in the hexagon case. Curves of different colors represent different vortices on cutting plane z=0.5 (except the dashed curves). The values are averaged over 10 planetary rotation periods toward the end of the simulation. (A) Mean tangential velocity $V^{c}$ around an individual vortex, plotted against radius to the center of the respective vortex. The thin and thick dashed curves are for the cutting planes $z=0.04$ and $z=0.96$, respectively. (B) Mean curvature vorticity $V^c/r$. (C) Relative vorticity $V^c/r+{dV}^c/dr$. (D) Inertial stability factor $f+V^c/r+{dV}^c/dr$.\label{fig:f10}}
\end{figure}

\subsection{Stability}

As the overall structures of the vortices vary little vertically due to dominance of rotation (convective Rossby number is much smaller than one), one may consider the motions in the medium scale (vortex scale) as quasi-two-dimensional, so that some simple analysis can be applied to explore the essential factors of the dynamics.

First, let us examine how far a vortex can be considered as axisymmetric. Around a vortex center, one can draw concentric circles and consider flow velocity component tangential to the circles. The positive direction is chosen to be counterclockwise (cyclonic). Distributions of the value of the tangential velocity for PC0 and HC0 are shown Figs.~\ref{fig:f7}A and \ref{fig:f7}B, respectively (for an instance). The dashed circles in the figures mark the 1.5 unit distance from the centers of the central cyclones. One can see that the tangential flows are close to axisymmetric at least within this distance. The dark blue contours (forming 5 and 6 blobs in A and B) outside the dashed circles show the negative values contributed by the neighboring cyclones. In contrast to the counterclockwise circulations near the centers, their averaged effect is to produce a circulation in the clockwise direction (anticyclonic). The same procedure can be performed for the off-center vortices and the qualitative features are the same.

One can make the analysis more quantitative by decomposing the tangential velocity on a circle into modes with different azimuthal wave numbers. On a circle with given radius, we perform a Fourier transform on the tangential velocity and find the spectral power contained in the different modes. The axisymmetric mode corresponds to wave number 0; the non-axisymmetric modes include contributions from the spiral structures, small scale turbulence, and interference from neighboring vortices. The faction of power contained in the $m=0$ mode relative to the total power provides a measure of how axisymmetric the velocity field around the vortex is. Fig.~\ref{fig:f8} plots radial variations of the fraction of spectral power in the axisymmetric mode (on the plane $z=0.5$); panels A and B are for Cases A and B, respectively. The blue curves in both figures represent the central vortices, and the orange curves represent the off-center vortices PC2 and HC2, respectively.  The two blue curves in both panels and the orange curve in panel B are rather flat and stay above 70\% within the distance 1.5. The orange curve in A stays above 70\% within the distance 2 (as the vortices on the pentagon are larger). Within these distances, the axisymmetric mode overwhelmingly dominates. All curves drop to 0 at some distances beyond 2. Those are locations where the mean rotation on a circle turns from cyclonic to anticyclonic. Beyond these distances, the systematic motions of neighboring vortices dominate.

In Fig.~\ref{fig:f9}A (pentagon case) and Fig.~\ref{fig:f10}A (hexagon case) the mean tangential velocities versus circle radius $r$ are plotted for all vortices (at an instance of time). The blue curves represent the central vortices, PC0 and HC0, with horizontal cuts at $z=0.5$. Solid curves of other colors represent the off-center vortices at the same cutting level. The thin and thick dashed curves represent cuttings of the two central vortices at z=0.04 and z=0.96, respectively.

There are a number of things to observe. First, all the mean tangential velocities first rise to certain peaks, then drop and turn negative at distances corresponding to the locations where the power of the $m=0$ mode becomes 0 in Fig.~\ref{fig:f8}. This reflects the situation as discussed in the previous two paragraphs. Second, for the pentagon case (Fig.~\ref{fig:f9}A), the positive ranges of the plots pertaining to the non-central cyclone extend over a wider distance than the blue line (PC0) as the off-center vortices are larger. The cuts of PC0 and HC0 at $z=0.04$ (near the bottom of the CZ) do not differ much from the blue curve, showing that the vortical flows do not change much vertically inside the CZ. On the other hand, the tangential velocity distributions of PC0 and HC0 at $z=0.96$ (just above the stability interface) drop substantially (though the profile remains similar). The fast radiative loss in the optically thin layer reduces the temperature and pressure contrasts between the inside and outside of the vortices, resulting in a reduction of vorticity (geostrophic and cyclostrophic balance). Third, at the radial distance where the circulation velocity peaks (around $r=1$), the vortex Rossby numbers $V^{c}_{peak}/(2\Omega r_{peak})$ of PC0 and HC0 at this height are about 0.4. They are larger than the value 0.24 estimated by \citet{li2020modeling} according to observation \citep{grassi2018first}.

Given that the vortex flow is almost axisymmetric up to a certain known radial distance, we can proceed to perform a stability analysis within the distance, based on analogy with an idealized axisymmetric flow having the same circulation velocity distribution. An inertial stability \citep{yanai1964formation,schubert1982inertial} criterion in rather general settings (stable and unstable backgrounds) has been given by \citet{kloosterziel2007inertial}. Here we focus on a convectively unstable background for which a \it necessary condition\rm\ for inertial stability is $\left(f+2V^c/r\right)\left(f+V^c/r+{dV}^c/dr\right)>0$ (see Appendix~\ref{appendixc}), where $r$ is the distance to the center of the vortex, $V^c$ is the axisymmetric flow velocity around the center, $f$ is the Coriolis parameter (or planetary vorticity), $V^c/r$ is the curvature vorticity (or mean relative angular velocity), and ${dV}^c/dr$ is the shear vorticity. For a system of multiple vortices, this linear stability criterion can be applied to flows around each of the vortices. If any of them violates the criterion, the system cannot be maintained. The combined requirement becomes a necessary condition for stability of the system.

We take Case A as example here. The mean tangential velocity $V^c$ versus $r$ is plotted for all the cyclones in Fig.~\ref{fig:f9}A (on the $z=0.5$ level). $V^c$ is always positive in the near field. At larger distances, $V^c$ turns negative as the circle for picking up the axisymmetric part of the circulation velocity expands into neighboring vortices (Fig.~\ref{fig:f7}A). The first term enclosed in parentheses on the left-hand-side of the stability inequality is always positive as the magnitude of $V^c/r$ is much less than $f$($\sim 1$\ for\ the\ current\ case)\ for large $r$ (Fig.~\ref{fig:f9}B). The second term enclosed in parentheses, however, contains consequential negative terms created by close packing of the vortices, but inertial stability requires positivity of this factor. The shear vorticity ${dV}^c/dr$ makes the greatest negative contribution to the second term enclosed in parentheses. The radial drop of $V^c$ is expedited by the shortened separations among the neighboring vortices. The tighter the packing, the more negative ${dV}^c/dr$ may reach. Despite the positivity of the curvature vorticity, the sum $V^c/r+{dV}^c/dr$ is negative beyond $r=1$ for PC0 and $r=1.8$ for PC1 to PC5 (Fig.~\ref{fig:f9}C). These radial distances are within the ranges where the vortex can safely be regarded as axisymmetric (see Fig.~\ref{fig:f8}). $f$ is the only term to keep the stability factor $f+V^c/r+{dV}^c/dr$ everywhere positive (Fig.~\ref{fig:f9}D). Thus it plays a critical role in allowing the cyclones to stay closely packed. A corollary is that cyclones cannot be packed closely for long in low latitudes. The analysis is similar for the hexagon case (see Fig.~\ref{fig:f10}).

Recently, \citet{li2020modeling} found that in shallow water formulation anticyclonic `rings' need to be introduced around individual cyclones to shield the cyclones from merging. This conclusion is compatible with the need for a significant Coriolis parameter. The rings cannot be stable without a large enough \it f\rm; that was implicitly satisfied by the selected vortex profiles. Around the computed vortices of our cases, anticyclonic `rings' do appear in the averaged sense (see Figs. 7, 9C, and 10C), and they arise naturally as a result of packing the vortices closely.

In the above discussion, the stability criterion is applied to each vortex separately. Global stability analysis of vortex polygons has been made by \citet{reinaud2019three}. Based on the quasi-geostrophic (QG) and inviscid assumptions, the author has shown that three-dimensional like-sign uniform PV (potential vorticity) vortices can remain stable in a m-fold axially symmetric configuration (with a moderate strength central vortex) for $m<7$.  The stable polygon patterns obtained by our simulations are in line with his conclusion. In contrast with the QG theory, our simulation includes viscous, nonadiabatic, turbulent effects, and the vortices are generated by the rotating turbulent convection.  Under the more complicated situation, the maintenance of closely packed vortices is more challenging because of the following two factors. First, the vortices could be destroyed by energy arising from viscous and turbulent dissipation. Second, turbulence could also enhance vortex merging so that the polygon pattern would be quickly destroyed. Our calculation has demonstrated that under favorable conditions (requirements set by the dimensionless flow parameters) multiple long-lived cyclones can be created and self-organized to a polygonal distribution at the polar region of a fast rotating, turbulent convection zone. It offers a viable mechanism to explain the long-lived closely-packed vortices observed in the polar regions of Jupiter.

\begin{figure}[!htbp]
\epsscale{0.9}
\plotone{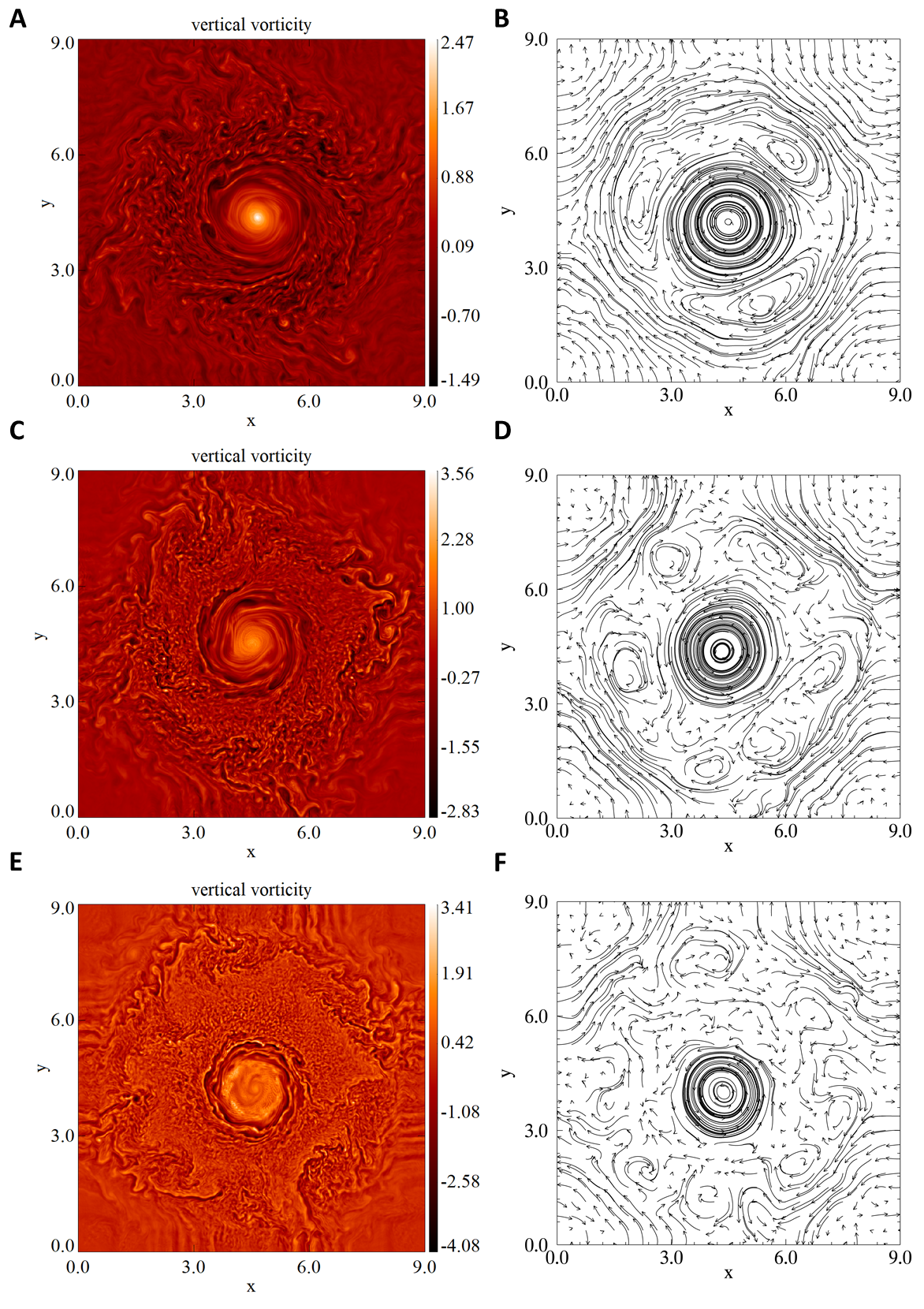}
\caption{Vertical vorticity (left-side panels) and horizontal velocity streamlines (right-side panel) at $z=0.85$ for simulation Cases C (first row), D (second row), and E (third row). For comparison with Saturn, one distance unit is about 8998 {\rm km}.\label{fig:f11}}
\end{figure}

\begin{deluxetable*}{ccccccccccccccccc}[htb!]
\tablecaption{Parameters of simulation cases for Saturn-like polar vortex\label{table:tab2}}
\tablehead{
 Case & $Ek$ & $Re$ & $Pr$ & $Ro^{c}$ & $F_{bot}$ & $\theta_{max}$ & $H/R$ & $N_{x}\times N_{y}\times N_{z}$
}
\startdata
C &   $1.1\times 10^{-5}$   & $7.3\times 10^3$ & 11.5 & 0.029 & 0.0025 & $40^{\circ}$ & 15.51\% & $900^2\times 101$  \\
D &   $5.6\times 10^{-6}$   & $6.2\times 10^3$ & 11.5 & 0.014 & 0.0025 & $40^{\circ}$ & 15.51\% & $900^2\times 101$   \\
E &   $3.6\times 10^{-6}$   & $5.1\times 10^3$ & 11.2 & 0.010 & 0.0025 & $40^{\circ}$ & 15.51\% & $900^2\times 101$   \\
\enddata
\tablecomments{The symbols have the same meanings as those in Table~\ref{table:tab1}. Each case has been run for over 32000 units of time (over 5000 planetary rotation periods).}
\end{deluxetable*}

\section{Discussion}\label{sec:discussion}

Although the focus of this paper is on Jupiter, the question about comparison with Saturn, which also has a deep convection zone, invariably arises. For deep convection models, the primary factor for the appearance of large-scale vortices is the convective Rossby number. The internal heat flux of Saturn is about one third of that of Jupiter. As convective velocity in a non-rotating CZ is approximately proportional to the cubic root of the convective energy flux \citep{bohm1958wasserstoffkonvektionszone,chan1989turbulent}, the velocity scale for calculating Saturn's $Ro^c$ would be roughly 0.7 times that of Jupiter. While the rotation period of Saturn is almost the same as that of Jupiter, Saturn's CZ is estimated to be about three times deeper \citep{iess2019measurement}. Therefore, the $Ro^c$ of Saturn can be estimated as $\sim 0.2$ times that of Jupiter, which means that the relative influence of rotation on Saturn is even stronger. Why did observations \citep{godfrey1988hexagonal,vasavada2006cassini,sanchez2014long} find only single long-lived cyclones in Saturn's polar regions?

This is an important question relevant to deciphering the vortex generation mechanism. Based on models with thin, stable layer settings, lower Rossby number should create more cyclones.
But in deep convection models, the thickness of the CZ is non-negligible and relevant. Besides the convective Rossby number, the depth of the convection zone also plays an important role in vortex formation (correspondingly the aspect ratio in an experimental setup) . If the depth of a $\gamma$-box increases, the aspect ratio decreases (the horizontal extent of the box has to stay within a reasonably range of colatitude from the pole). A box with small aspect ratio cannot accommodate many vortices. Though dependent on the Rossby number, the intrinsic aspect ratio (diameter over depth) of a deep vortex in the polar region, at least for the range of $Ro^c$ reachable so far, is on the order of 3 (see for example Fig.~\ref{fig:f3}). As the depth of Saturn's convection zone is estimated to be three times deeper than that of Jupiter, we expect that the aspect ratio of Saturn's polar convection zone is about one third of that of Jupiter. Its polar vortex is thus much larger and the number is limited to 1. This argument ties the number of long-lived vortices directly with the depth of the convection zone relative to the planetary radius (the parameter $H/R$).

To examine the argument above, we have performed some preliminary numerical experiments for Saturn-like configurations. Table~\ref{table:tab2} lists the parameters of three additional simulation cases (labeled by C, D, and E). The particular characteristics of these cases are the three-fold increase of $H/R$, the doubling of $\theta_{max}$, and the successive decrease of  $Ro^c$ (the minimum value is four times lower than that of Table~\ref{table:tab1}). The number of horizontal grids is smaller, as the aspect ratio of the computational box is smaller (by a factor of 9/16 relative to that of Cases A and B; also see Appendix B). The left-side panels of Fig.~\ref{fig:f11} show examples of horizontal cuts of the vertical vorticity at $z=0.85$ (the stability interface is now at $z=0.75$) for the three cases C, D, and E in sequential order; the right-side panels of the same figure show corresponding horizontal velocity streamlines. The streamlines of Case C (Fig.~\ref{fig:f11}B) indicate some similarity with the polar vortex obtained recently by the model of \citet{yadav2020deep}, which was an anelastic global simulation of deep rotating convection ($H/R=0.1$, 5 density scale heights). They found a single large cyclone at the pole surrounded by a band of three large anticyclones. Our cases show that as $Ro^c$ decreases, more anticyclones appear in the surrounding band and polygonal features may arise in flows around the polar vortex, within which the distribution of positive PV broadens. The associated phenomena are intriguing but we postpone an investigation to later study.

\section{Conclusion}
In summary, we have shown that numerical computation solving the full set of hydrodynamics equation for rapidly rotating deep convection could produce both Jupiter-like closely-packed circumpolar cyclones and Saturn-like solitary polar cyclone. The convective Rossby number and the lateral-to-height aspect ratio are confirmed to play critical roles in the formation and the selection of the number of these polar cyclones. For each of the cyclones in the Jovian cases, we analysed the axisymmetric component of the tangential velocity field around the vortex and found that a large enough Coriolis parameter is a necessary condition for allowing closely-packed, multiple cyclones to coexist.

Due to the limitation of computer resources, we only give a few examples to illustrate the physical processes of how deep polar vortices on Jupiter and Saturn may be formed and maintained. Quantitative comparisons with observations have shown that these models are generally inadequate.  One major reason could be the very large energy fluxes introduced for fast relaxation (so that velocities become higher and changes are faster). Simulations with more physical realism need more powerful codes (e.g. \citet{cai2016semi}); they will be very useful for closer comparison with observational results.

\appendix
\section{Dimensional Quantities}\label{appendixa}
As the model is based on similarity of dimensionless parameters, the  quantities are made dimensionless. We scale all the physical variables by the pressure, density, temperature, and height at the top of the box. In Table~\ref{table:tab3} $p_{scl}$, $\rho_{scl}$, $T_{scl}$, and $L_{scl}$ represent the scales of pressure, density, temperature, and length, respectively. Consequently, the velocity scale is $v_{scl}=(p_{scl}/\rho_{scl})^{1/2}$; the time scale is $t_{scl}=L_{scl}/v_{scl}$; and the flux scale is $F_{scl}=\rho_{scl} v_{scl}^3$.
For comparison with dimensional quantities of Jupiter and Saturn, the values of the essential scale factors are listed.  A dimensional quantity can be obtained by multiplying the dimensionless value with the corresponding scale factor.

\begin{deluxetable*}{ccccccccccccccccc}[!htbp]
\tablecaption{Dimensional scales for Jupiter and Saturn \label{table:tab3}}
\tablehead{
 Case & $L_{scl}$ & $p_{scl}$ & $\rho_{scl}$ & $T_{scl}$ & $v_{scl}$ & $t_{scl}$ & $F_{scl}$ & planet
}
\startdata
A &  3668 {\rm km}   &  $10^5$ {\rm Pa} & 0.167 $\rm kg/m^3$ & 166 {\rm K} & 773.8 {\rm m/s} & 4740 {\rm s} & $7.73\times 10^{7} {\rm W/m^2}$ & Jupiter-like  \\
B &  1841 {\rm km}   &  $10^5$ {\rm Pa} & 0.167 $\rm kg/m^3$ & 166 {\rm K} & 773.8 {\rm m/s} & 2379 {\rm s} & $7.73\times 10^{7} {\rm W/m^2}$ & Jupiter-like  \\
C-E &  8998 {\rm km}   &  $10^5$ {\rm Pa} & 0.199 $\rm kg/m^3$ & 137 {\rm K} & 708.9 {\rm m/s} & 8998 {\rm s} & $7.09\times 10^{7} {\rm W/m^2}$ & Saturn-like   \\
\enddata
\end{deluxetable*}

\section{Influence of grid resolution}\label{appendixb}
\begin{figure}[!htbp]
\plotone{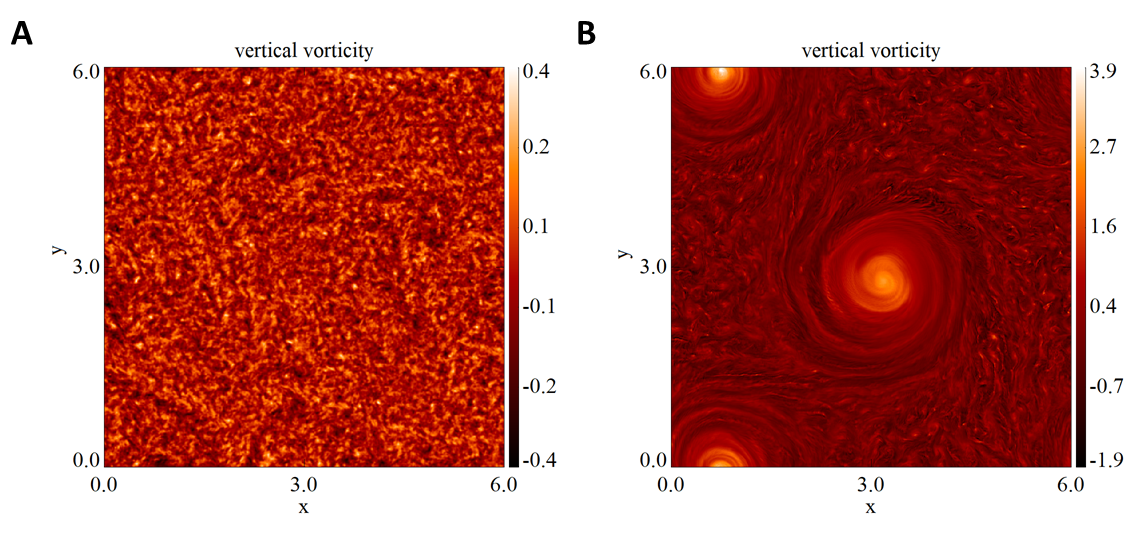}
\caption{Comparison of vertical vorticity distributions on the $z=0.75$ cutting plane. (A) The low-resolution Case R1; (B) The high-resolution Case R2. \label{fig:f12}}
\end{figure}

To illustrate the effect of grid resolution on vortex formation, we show here an example with two simulations, Cases R1 and R2, in a horizontally periodic polar {\it f}-plane box configuration (constant vertical $\bf\Omega$) with aspect ratio $6$. The parameters of the two cases are identical except the resolution. The convective Rossby number $Ro^c$ is equal to 0.059. A resolution of $N_{x}\times N_{y}\times N_{z}=450^2\times 75$ is used in Case R1, while a higher resolution of $900^2\times 149$ is used in Case R2. Both cases were initially perturbed with the same small random velocity field and then evolved for about 750 planetary rotation periods. Their final structures as described by the vertical component of vorticity on the plane $z=0.75$ are shown in Figs.~\ref{fig:f12}A and ~\ref{fig:f12}B. A pair of cyclones appear in the finer-grid case, while no large-scale cyclone is found in the coarser-grid case. Besides illustrating the importance of resolution, this comparison also shows that the horizontal resolution of $900^2$ is adequate to accommodate more than one roaming cyclone in a polar {\it f}-plane box with moderate aspect ratio, when there is no polar beta effect to push them to merge at the pole (as in Cases C-E).

\section{Inertial stability in convectively unstable situation}\label{appendixc}
Under the Boussinesq approximation, it can be proved that the following condition \citep{kloosterziel2007inertial} should be satisfied for an axisymmetric flow:
\begin{eqnarray}
\int_{\mathcal{V}}\left[(\omega^2+N^2)\left|\frac{1}{r}\frac{\partial \Psi r}{\partial r}\right|^2+(\omega^2+\Phi)\left|\frac{\partial \Psi}{\partial z}\right|^2\right]d\mathcal{V}=0,
\end{eqnarray}
where $\psi=\exp(\omega t)\Psi(r,z)$ is the meridional streamfunction of the perturbed velocity, $\omega$ is the time frequency, $r$ is the radius away from the vortex center, $z$ is the height, $N^2$ is the Brunt V\"ais\"al\"a frequency,  $\Phi=(f+2V^{c}/r)(f+V^{c}/r+dV^{c}/dr)$ is the Rayleigh discriminant, and $\mathcal{V}$ is the volume of the domain of $\Psi$. With some rearrangement, the time frequency square can be expressed as
\begin{eqnarray}
\omega^2=-\frac{I_{1}}{I_{2}}~,
\end{eqnarray}
where
\begin{eqnarray}
I_{1}=\int_{\mathcal{V}} \left[N^2\left|\frac{1}{r}\frac{\partial \Psi r}{\partial r}\right|^2 +\Phi\left|\frac{\partial \Psi}{\partial z}\right|^2\right] d\mathcal{V}~,
\end{eqnarray}
and
\begin{eqnarray}
I_{2}=\int_{\mathcal{V}} \left[\left|\frac{1}{r}\frac{\partial \Psi r}{\partial r}\right|^2 +\left|\frac{\partial \Psi}{\partial z}\right|^2\right] d\mathcal{V}~.
\end{eqnarray}
The flow is stable to axisymmetric modes when $\omega^2<0$. As $I_{2}$ is positive definite, a stable vortex requires $I_{1}>0$. The sign of $I_{1}$ depends on $N^2$ and $\Phi$. If $N^2>0$, then $\Phi>0$ is a sufficient but not a necessary condition for inertial stability (under axisymmetric perturbations). If $N^2=0$, then $\Phi>0$ is a sufficient and necessary condition. If $N^2<0$, then $\Phi>0$ is \it a necessary but not sufficient condition \rm for inertial stability. Though our fluid is non-Boussinesq, this criterion provides a helpful analytical framework for theoretical understanding.

\section{Supplementary movies}\label{appendixd}
\begin{figure}[!htbp]
    \begin{interactive}{animation}{MovieS1.mp4}
    \plotone{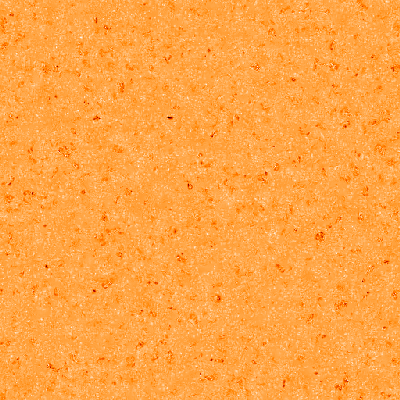}
    \end{interactive}
    \caption{Static figure for MovieS1. This movie shows the development of small vortices in a small {\it f}-box, as a preparation phase of Case B. The aspect ratio of the box is 4. The time period of this movie is about 1332 units of time (about 98 planetary rotation periods). The bright (dark) color represents higher (lower) temperature. \label{fig:movies1}}
\end{figure}

\begin{figure}[!htbp]
    \begin{interactive}{animation}{MovieS2.mp4}
    \plotone{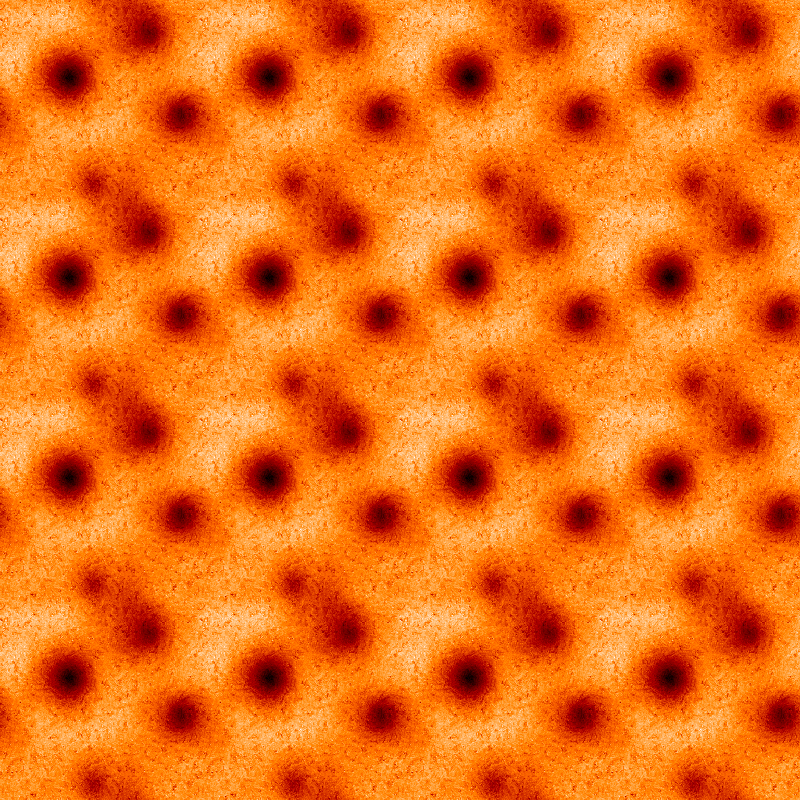}
    \end{interactive}
    \caption{Static figure for MovieS2.  This movie shows the merger of small vortices, growth and saturation of vortex sizes, and formation of the hexagonal pattern. The aspect ratio of the box is 16. The time period of this movie is over 33000 units of time (over 2600 planetary rotation periods). The bright (dark) color represents higher (lower) temperature.\label{fig:movies2}}
\end{figure}

\begin{figure}[!htbp]
    \begin{interactive}{animation}{MovieS3.mp4}
    \plotone{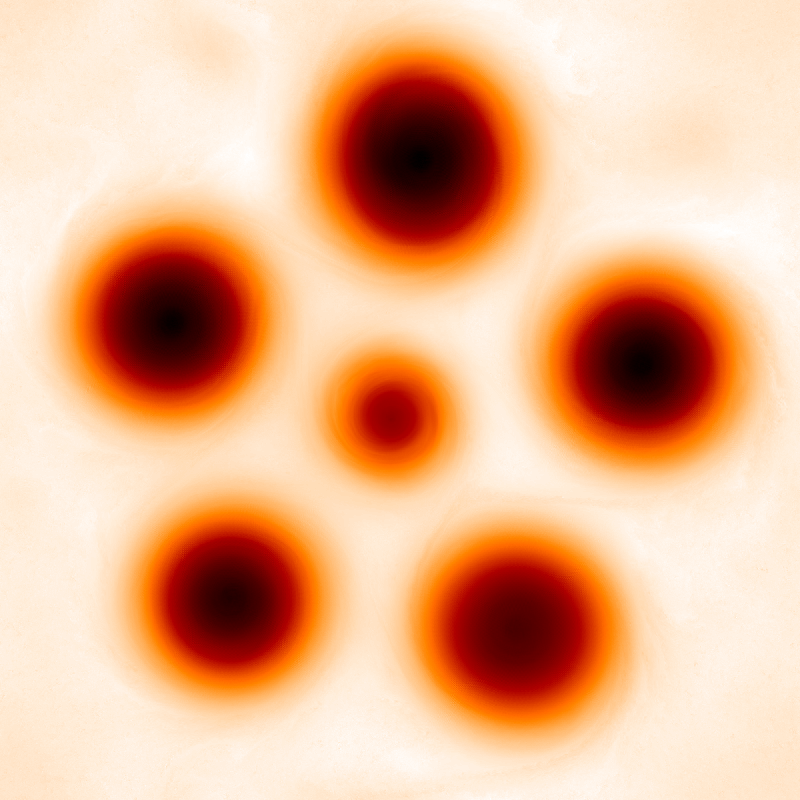}
    \end{interactive}
    \caption{Static figure for MovieS3. This movie shows the time evolution and persistence of the pentagonal pattern. The time period of this movie is about 1200 units of time (around 96 planetary rotation periods). The bright (dark) color represents higher (lower) temperature. \label{fig:movies3}}
\end{figure}

MovieS1 (Fig.~\ref{fig:movies1} is the static figure) shows the generation of small vortices in a small {\it f}-box. The aspect ratio of the box is 4. The time period of this movie is 1332 (about 98 planetary rotation periods). MovieS2 (Fig.~\ref{fig:movies2} is the static figure) shows the generation and formation of the hexagon pattern by the closely-packed cyclones. The aspect ratio of the box is 16. The time period of this movie is over 33000 (over 2600 planetary rotation periods). MovieS3 (Fig.~\ref{fig:movies3} is the static figure) shows the time evolution of the pentagon pattern. The time period of this movie is 1200 (around 96 planetary rotation periods). In all the movies, the bright (dark) color represents higher (lower) temperature. The horizontal cuts of MovieS1 and MovieS2 are at $z=0.29$ and that of MovieS3 is at $z=0.22$.

\section*{Acknowledgements}
We are grateful to the two anonymous reviewers for their comments and suggestions that have led to major improvement of the manuscript. We thank Alessandro Mura, Alberto Adriani, and the Juno team for kindly providing us with their JIRAM image (Fig.~\ref{fig:f1}C). We thank Yamila Miguel and Tristan Guillot for kindly providing us with their latest result on the interior structure of Saturn. The physical values at the surface of Saturn are taken from their model result. We also thank Dongdong Ni for discussion on the interior structure of Jupiter. Part of the simulation was performed on the supercomputers at the China National Supercomputer Center in Guangzhou. This work was partially supported by the Science and Technology Development Fund, Macau SAR (Nos.0045/2018/AFJ, 119/2017/A3, 0156/2019/A3), NSFC (Nos.11503097, 11521101), the Guangdong Basic and Applied Basic Research Foundation (No.2019A1515011625), the Science and Technology Program of Guangzhou (No.201707010006), and the Hong Kong Research Grants Council (HKUST 600309).




\bibliographystyle{aasjournal}
\bibliography{main}

\begin{thebibliography}{}
\expandafter\ifx\csname natexlab\endcsname\relax\def\natexlab#1{#1}\fi
\providecommand{\url}[1]{\href{#1}{#1}}
\providecommand{\dodoi}[1]{doi:~\href{http://doi.org/#1}{\nolinkurl{#1}}}
\providecommand{\doeprint}[1]{\href{http://ascl.net/#1}{\nolinkurl{http://ascl.net/#1}}}
\providecommand{\doarXiv}[1]{\href{https://arxiv.org/abs/#1}{\nolinkurl{https://arxiv.org/abs/#1}}}

\bibitem[{Adriani {et~al.}(2018)Adriani, Mura, Orton, Hansen, Altieri,
  Moriconi, Rogers, Eichst{\"a}dt, Momary, Ingersoll,
  {et~al.}}]{adriani2018clusters}
Adriani, A., Mura, A., Orton, G., {et~al.} 2018, Nature, 555, 216

\bibitem[{Adriani {et~al.}(2020)Adriani, Bracco, Grassi, Moriconi, Mura, Orton,
  Altieri, Ingersoll, Atreya, Lunine, {et~al.}}]{adriani2020two}
Adriani, A., Bracco, A., Grassi, D., {et~al.} 2020, Journal of Geophysical
  Research: Planets, e2019JE006098

\bibitem[{Aurnou {et~al.}(2015)Aurnou, Calkins, Cheng, Julien, King, Nieves,
  Soderlund, \& Stellmach}]{aurnou2015rotating}
Aurnou, J., Calkins, M., Cheng, J., {et~al.} 2015, Physics of the Earth and
  Planetary Interiors, 246, 52

\bibitem[{Bagenal {et~al.}(2007)Bagenal, Dowling, \&
  McKinnon}]{bagenal2007jupiter}
Bagenal, F., Dowling, T.~E., \& McKinnon, W.~B. 2007, Jupiter: the planet,
  satellites and magnetosphere, Vol.~1 (Cambridge University Press)

\bibitem[{B{\"o}hm-Vitense(1958)}]{bohm1958wasserstoffkonvektionszone}
B{\"o}hm-Vitense, E. 1958, Zeitschrift fur Astrophysik, 46, 108

\bibitem[{Bolton {et~al.}(2017)Bolton, Adriani, Adumitroaie, Allison, Anderson,
  Atreya, Bloxham, Brown, Connerney, DeJong, {et~al.}}]{bolton2017jupiter}
Bolton, S.~J., Adriani, A., Adumitroaie, V., {et~al.} 2017, Science, 356, 821

\bibitem[{Brueshaber {et~al.}(2019)Brueshaber, Sayanagi, \&
  Dowling}]{brueshaber2019dynamical}
Brueshaber, S.~R., Sayanagi, K.~M., \& Dowling, T.~E. 2019, Icarus, 323, 46

\bibitem[{Cai(2016)}]{cai2016semi}
Cai, T. 2016, Journal of Computational Physics, 310, 342

\bibitem[{Cerretelli \& Williamson(2003)}]{cerretelli2003physical}
Cerretelli, C., \& Williamson, C. 2003, Journal of Fluid Mechanics, 475, 41

\bibitem[{Chan(2007)}]{chan2007rotating}
Chan, K.~L. 2007, Astronomische Nachrichten: Astronomical Notes, 328, 1059

\bibitem[{Chan \& Mayr(2013)}]{chan2013numerical}
Chan, K.~L., \& Mayr, H.~G. 2013, Earth and Planetary Science Letters, 371, 212

\bibitem[{Chan \& Sofia(1986)}]{chan1986turbulent}
Chan, K.~L., \& Sofia, S. 1986, The Astrophysical Journal, 307, 222

\bibitem[{Chan \& Sofia(1989)}]{chan1989turbulent}
---. 1989, The Astrophysical Journal, 336, 1022

\bibitem[{Favier {et~al.}(2014)Favier, Silvers, \& Proctor}]{favier2014inverse}
Favier, B., Silvers, L., \& Proctor, M. 2014, Physics of Fluids, 26, 096605

\bibitem[{Folz \& Nomura(2017)}]{folz2017quantitative}
Folz, P.~J., \& Nomura, K.~K. 2017, Journal of Fluid Mechanics, 829, 1

\bibitem[{Godfrey(1988)}]{godfrey1988hexagonal}
Godfrey, D. 1988, Icarus, 76, 335

\bibitem[{Grassi {et~al.}(2018)Grassi, Adriani, Moriconi, Mura,
  Tabataba-Vakili, Ingersoll, Orton, Hansen, Altieri, Filacchione,
  {et~al.}}]{grassi2018first}
Grassi, D., Adriani, A., Moriconi, M., {et~al.} 2018, Journal of Geophysical
  Research: Planets, 123, 1511

\bibitem[{Guervilly \& Hughes(2017)}]{guervilly2017jets}
Guervilly, C., \& Hughes, D.~W. 2017, Physical Review Fluids, 2, 113503

\bibitem[{Guervilly {et~al.}(2014)Guervilly, Hughes, \&
  Jones}]{guervilly2014large}
Guervilly, C., Hughes, D.~W., \& Jones, C.~A. 2014, Journal of Fluid Mechanics,
  758, 407

\bibitem[{Guzm{\'a}n {et~al.}(2020)Guzm{\'a}n, Madonia, Cheng,
  Ostilla-M{\'o}nico, Clercx, \& Kunnen}]{guzman2020competition}
Guzm{\'a}n, A. J.~A., Madonia, M., Cheng, J.~S., {et~al.} 2020, Physical Review
  Letters, 125, 214501

\bibitem[{Hanel {et~al.}(1981)Hanel, Conrath, Herath, Kunde, \&
  Pirraglia}]{hanel1981albedo}
Hanel, R., Conrath, B., Herath, L., Kunde, V., \& Pirraglia, J. 1981, Journal
  of Geophysical Research: Space Physics, 86, 8705

\bibitem[{Heimpel {et~al.}(2016)Heimpel, Gastine, \&
  Wicht}]{heimpel2016simulation}
Heimpel, M., Gastine, T., \& Wicht, J. 2016, Nature Geoscience, 9, 19

\bibitem[{Horedt(2004)}]{horedt2004polytropes}
Horedt, G.~P. 2004, Polytropes: applications in astrophysics and related
  fields, Vol. 306 (Springer Science \& Business Media)

\bibitem[{Hurlburt {et~al.}(1984)Hurlburt, Toomre, \&
  Massaguer}]{hurlburt1984two}
Hurlburt, N., Toomre, J., \& Massaguer, J. 1984, The Astrophysical Journal,
  282, 557

\bibitem[{Iess {et~al.}(2019)Iess, Militzer, Kaspi, Nicholson, Durante,
  Racioppa, Anabtawi, Galanti, Hubbard, Mariani,
  {et~al.}}]{iess2019measurement}
Iess, L., Militzer, B., Kaspi, Y., {et~al.} 2019, Science, 364

\bibitem[{K{\"a}pyl{\"a} {et~al.}(2011)K{\"a}pyl{\"a}, Mantere, \&
  Hackman}]{kapyla2011starspots}
K{\"a}pyl{\"a}, P.~J., Mantere, M.~J., \& Hackman, T. 2011, The Astrophysical
  Journal, 742, 34

\bibitem[{Kaspi {et~al.}(2018)Kaspi, Galanti, Hubbard, Stevenson, Bolton, Iess,
  Guillot, Bloxham, Connerney, Cao, {et~al.}}]{kaspi2018jupiter}
Kaspi, Y., Galanti, E., Hubbard, W.~B., {et~al.} 2018, Nature, 555, 223

\bibitem[{Kloosterziel {et~al.}(2007)Kloosterziel, Carnevale, \&
  Orlandi}]{kloosterziel2007inertial}
Kloosterziel, R., Carnevale, G., \& Orlandi, P. 2007, Journal of Fluid
  Mechanics, 583, 379

\bibitem[{Li {et~al.}(2020)Li, Ingersoll, Klipfel, \& Brettle}]{li2020modeling}
Li, C., Ingersoll, A.~P., Klipfel, A.~P., \& Brettle, H. 2020, Proceedings of
  the National Academy of Sciences, 117, 24082

\bibitem[{Li {et~al.}(2018)Li, Jiang, West, Gierasch, Perez-Hoyos,
  Sanchez-Lavega, Fletcher, Fortney, Knowles, Porco, {et~al.}}]{li2018less}
Li, L., Jiang, X., West, R., {et~al.} 2018, Nature Communications, 9, 1

\bibitem[{Marcus(1990)}]{marcus1990vortex}
Marcus, P.~S. 1990, Journal of Fluid Mechanics, 215, 393

\bibitem[{Merlis {et~al.}(2016)Merlis, Zhou, Held, \& Zhao}]{merlis2016surface}
Merlis, T.~M., Zhou, W., Held, I.~M., \& Zhao, M. 2016, Geophysical Research
  Letters, 43, 2859

\bibitem[{Nof(1990)}]{nof1990modons}
Nof, D. 1990, Geophysical \& Astrophysical Fluid Dynamics, 52, 71

\bibitem[{O'Neill {et~al.}(2015)O'Neill, Emanuel, \& Flierl}]{o2015polar}
O'Neill, M.~E., Emanuel, K.~A., \& Flierl, G.~R. 2015, Nature Geoscience, 8,
  523

\bibitem[{O'Neill {et~al.}(2016)O'Neill, Emanuel, \& Flierl}]{o2016weak}
---. 2016, Journal of the Atmospheric Sciences, 73, 1841

\bibitem[{Reinaud(2019)}]{reinaud2019three}
Reinaud, J.~N. 2019, Journal of Fluid Mechanics, 863, 32

\bibitem[{S{\'a}nchez-Lavega {et~al.}(2014)S{\'a}nchez-Lavega, del
  R{\'\i}o-Gaztelurrutia, Hueso, P{\'e}rez-Hoyos, Garc{\'\i}a-Melendo,
  Antu{\~n}ano, Mendikoa, Rojas, Lillo, Barrado-Navascu{\'e}s,
  {et~al.}}]{sanchez2014long}
S{\'a}nchez-Lavega, A., del R{\'\i}o-Gaztelurrutia, T., Hueso, R., {et~al.}
  2014, Geophysical Research Letters, 41, 1425

\bibitem[{Schecter \& Dubin(1999)}]{schecter1999vortex}
Schecter, D.~A., \& Dubin, D.~H. 1999, Physical Review Letters, 83, 2191

\bibitem[{Schubert \& Hack(1982)}]{schubert1982inertial}
Schubert, W.~H., \& Hack, J.~J. 1982, Journal of the Atmospheric Sciences, 39,
  1687

\bibitem[{Scott(2011)}]{scott2011polar}
Scott, R. 2011, Geophysical \& Astrophysical Fluid Dynamics, 105, 409

\bibitem[{Smagorinsky(1963)}]{smagorinsky1963general}
Smagorinsky, J. 1963, Monthly Weather Review, 91, 99

\bibitem[{Stellmach {et~al.}(2014)Stellmach, Lischper, Julien, Vasil, Cheng,
  Ribeiro, King, \& Aurnou}]{stellmach2014approaching}
Stellmach, S., Lischper, M., Julien, K., {et~al.} 2014, Physical Review
  Letters, 113, 254501

\bibitem[{Tabataba-Vakili {et~al.}(2020)Tabataba-Vakili, Rogers, Eichst{\"a}dt,
  Orton, Hansen, Momary, Sinclair, Giles, Caplinger, Ravine,
  {et~al.}}]{tabataba2020long}
Tabataba-Vakili, F., Rogers, J., Eichst{\"a}dt, G., {et~al.} 2020, Icarus, 335,
  113405

\bibitem[{Vasavada {et~al.}(2006)Vasavada, H{\"o}rst, Kennedy, Ingersoll,
  Porco, Del~Genio, \& West}]{vasavada2006cassini}
Vasavada, A.~R., H{\"o}rst, S., Kennedy, M., {et~al.} 2006, Journal of
  Geophysical Research: Planets, 111

\bibitem[{Yadav \& Bloxham(2020)}]{yadav2020deep}
Yadav, R.~K., \& Bloxham, J. 2020, Proceedings of the National Academy of
  Sciences

\bibitem[{Yanai(1964)}]{yanai1964formation}
Yanai, M. 1964, Reviews of Geophysics, 2, 367

\end{thebibliography}



\end{document}